\documentclass[11pt,a4paper]{article}
\pdfoutput=1
\usepackage[english]{babel}
\usepackage{amssymb,amsmath,latexsym,mathrsfs}
\usepackage{feynmp-auto} 
\usepackage{graphicx}
\usepackage{makeidx}
\usepackage{epsfig}
\usepackage{multirow}
\usepackage{array}
\usepackage{color}
\usepackage{jcappub}
\usepackage{longtable}
\usepackage{multirow}
\usepackage{bigstrut}
\usepackage{hyperref}

\newcommand{\MeV}{\mathrm{MeV}}

\newcommand{\Mpc}{\mathrm{Mpc}}
\newcommand{\neff}{N_\mathrm{eff}}
\newcommand{\gnn}{\gamma_{\nu\nu}}
\newcommand{\zrec}{z_{\nu\mathrm{rec}}}
\newcommand{\tauc}{\tau_\mathrm{c}}
\newcommand{\taureio}{\tau_\mathrm{rec}}

\title{Constraints on secret neutrino interactions after Planck}
\author[a,b]{Francesco Forastieri}
\author[a,b]{, Massimiliano Lattanzi}
\author[a,b,c]{and Paolo Natoli}
\affiliation[a]{Dipartimento di Fisica e Scienze della Terra, Universit\`a degli Studi di Ferrara, via Giuseppe Saragat 1, I-44122 Ferrara, Italy}
\affiliation[b]{Istituto Nazionale di Fisica Nucleare, Sezione di Ferrara, via Giuseppe Saragat 1, I-44122 Ferrara, Italy}
\affiliation[c]{Agenzia Spaziale Italiana Science Data Center, Via del Politecnico
snc, 00133, Roma, Italy}
\emailAdd{francesco.forastieri@unife.it}
\emailAdd{lattanzi@fe.infn.it}
\emailAdd{natoli@fe.infn.it}

\abstract{
Neutrino interactions beyond the standard model of particle physics may affect
the cosmological evolution and can be constrained through observations.
We consider the possibility that neutrinos possess secret scalar or pseudoscalar
interactions mediated by the Nambu-Goldstone boson of a still unknown spontaneously 
broken global $U(1)$ symmetry, as in, \emph{e.g.}, Majoron models. In such scenarios, neutrinos
still decouple at $T\simeq 1\,\MeV$, but become tightly coupled again (``recouple'')
at later stages of the cosmological evolution. We use available 
observations of the cosmic microwave background (CMB) anisotropies, including Planck 2013 and the joint
BICEP2/Planck 2015 data,
to derive constraints on the quantity $\gnn^4$, parameterizing the neutrino collision rate
due to scalar or pseudoscalar interactions. We consider both a minimal
extension of the standard $\Lambda$CDM model, and more complicated scenarios with extra
relativistic degrees of freedom or non-vanishing tensor amplitude. For a wide range of dataset and model
combinations, we find a typical constraint $\gnn^4 \lesssim 0.9\times 10^{-27}$ (95\% C.L.),
implying an upper limit on the redshift $\zrec$ of neutrino recoupling $\lesssim 8500$, leaving
open the possibility that the latter occured well before hydrogen recombination. In the framework of Majoron models, the
upper limit on $\gnn$ roughly translates on a constraint $g \lesssim 8.2\times 10^{-7}$ on 
the Majoron-neutrino coupling constant $g$.
In general, the data show a weak ($\sim 1\sigma$) but intriguing preference for non-zero values of $\gnn^4$, with best fits 
in the range $\gnn^4 = (0.15 - 0.35)\times 10^{-27}$, depending on the particular dataset. This is more evident when either high-resolution 
CMB observations from the ACT and SPT experiments are included, 
or the possibility of non-vanishing tensor modes is considered. In particular, for the minimal model 
$\Lambda$CDM+$\gnn$ and including the Planck 2013, ACT and SPT data, we report $\gnn^4=(0.44^{+0.17}_{-0.36})\times10^{-27}$ 
($300 \lesssim \zrec \lesssim 5500$)  at 68\% confidence level.

}
\arxivnumber{}

\begin{document}
\maketitle

\section{Introduction}

Cosmological observations are a powerful probe of neutrino physics. To date,
all the available cosmological data are consistent with the expectation, based on the 
standard model of particle physics, that the Universe is filled with a thermal background of relic neutrinos
with $T_\nu = 1.9$~K, belonging to three families, each
contributing 113 particles per cm$^3$ to the total neutrino abundance.
In the early Universe, neutrinos are kept in thermal equilibrium with the cosmological fluid
by weak interactions until the expansion brings the temperature down to $T\simeq 1\,\MeV$. 
At later times, the interaction rate becomes too small and neutrinos decouple.
However, since the decoupling happens when neutrinos are ultrarelativistic, they keep a thermal spectrum
whose temperature scales as the inverse of the cosmological scale factor $a$. Thus, in the 
standard cosmological model (SCM), the only free parameters the `neutrino sector' of the model 
are the masses of the three eigenstates. Oscillation experiments have measured mass differences, showing that
at least two of the eigenstates have non-vanishing masses. However, both the absolute scale and the 
hierarchy of the masses remain unknown. Present-day cosmological observations can tightly constrain 
the sum of the masses \cite{sergio,sergio2}: Planck measurements
of the temperature and polarization anisotropies of the cosmic microwave background (CMB),
when combined with other astrophysical datasets, imply $\sum{m_\nu} < 0.23$ \cite{Planck:2015xua}.

This simple picture, other than being theoretically well-grounded, is perfectly consistent with all the available data.
It is however desirable to keep an open mind and test more complicated scenarios for the neutrino sector
of the SCM. For example, cosmological data can be used, among others, to constrain lepton asymmetry \cite{Castorina:2012md},
possible deviations of the neutrino spectrum from equilibrium \cite{Cuoco:2005qr}, and also the properties of a sterile neutrino 
with $\sim$ eV mass \cite{Melchiorri:2008gq,Archidiacono:2012ri,Archidiacono:2013xxa,Mirizzi:2013kva,DiValentino:2013qma,Giusarma:2014zza,Zhang:2014dxk,Dvorkin:2014lea,Archidiacono:2014apa,Zhang:2014nta,Leistedt:2014sia,Bergstrom:2014fqa,Costanzi:2014tna,Saviano:2014esa,Tang:2014yla}, that could possibly explain reactor anomalies (see e.g. Refs. \cite{Abazajian:2012ys,Kopp:2013vaa} for a review). In this paper, we will consider the possibility that
neutrinos have interactions beyond the standard model of particle physics (that for simplicity
we shall call ``hidden''  or ``secret'' interactions) and study the constraining power
of cosmological observations with respect to such a scenario.
In particular, we will consider a specific version of secret interactions, that is however
representative of a large class of models, namely a (pseudo)scalar interaction mediated
by the Nambu-Goldstone boson of a hitherto unknown broken $U(1)$ symmetry, like in Majoron models
\cite{Chikashige:1980ui,Schechter:1981cv,Gelmini:1980re}.
These models are very well-motivated  from the point of view of particle physics and have
the peculiar feature that the ratio between the interaction and expansion rates \emph{increases} with time,
so that neutrinos may actually become kinematically coupled again in the late stages of the cosmological evolution. 

Non-standard neutrino interactions have first been considered in a cosmological context
in Ref. \cite{Beacom:2004yd}, discussing the possibility that neutrino decay induced by the new interaction
would lead to a neutrinoless Universe. Limits on neutrino-neutrino scattering 
induced by non-standard interactions (either Majoron-like, as those considered in this paper, or Fermi-like) from
cosmological observations have been derived in Refs. \cite{Bell:2005dr,Basboll:2006yx,Cyr-Racine:2013jua} and, more recently,
in Ref. \cite{Archidiacono:2013dua}, using data from the Planck 2013 release. In these papers,
the neutrino fluid is modeled as abruptly changing from collisionless  to perfectly tightly coupled (or viceversa
in the case of Fermi-like interactions) at a given transition redshift, that represents the parameter actually
constrained by the data. A complementary approach 
consists in deriving limits on phenomenological quantities parameterizing the effective sound speed and viscosity
of the neutrino fluid \cite{Trotta:2004ty,Smith:2011es,Archidiacono:2012gv,Diamanti:2012tg,Archidiacono:2013lva,Gerbino:2013ova,Audren:2014lsa}.
As noted by a few authors, however, this approach does not always provide an accurate representation of the collisional regime 
\cite{Cyr-Racine:2013jua,Oldengott:2014qra}. For this reason we avoid resorting to it throughout this paper.
Instead, we derive limits on the strength of neutrino non-standard interactions
by directly modifying the Boltzmann equation in order to account for neutrino collisions, without
assuming a sudden transition between the two limiting regimes (free-streaming and tight coupling). We also
consider extended models allowing for extra relativistic species or tensor pertubations.

This paper is structured as follows. In Sec.~\ref{sec:theory} we briefly introduce the theoretical framework that
describes the hidden interactions of interest. In Sec. \ref{sec:boltz} we review the Boltzmann formalism
for interacting neutrinos. In Sec. \ref{sec:method} we describe the method used to compute 
the impact of interacting neutrinos on the evolution of cosmological perturbations and on the CMB observables, 
and to derive constraints on the strength of the interaction, that are discussed in Sec. \ref{sec:res}. Finally, in Sec \ref{sec:conc}
we draw our conclusions.

\section{Hidden neutrino interactions}
\label{sec:theory}

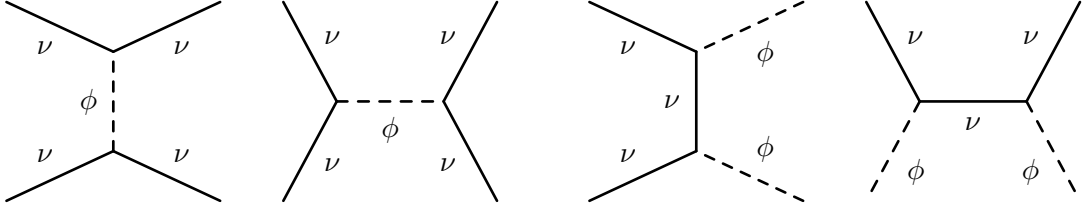
\begin{figure}[t]
\begin{fmffile}{feynman1}
\begin{fmfgraph*}(100,75)
\fmfleft{i1,i2}
\fmfright{o1,o2}
\fmf{plain, label=$\nu$}{v1,i1}
\fmf{plain, label=$\nu$}{o1,v1}
\fmf{plain, label=$\nu$}{i2,v2}
\fmf{plain, label=$\nu$}{o2,v2}
\fmf{dashes, label=$\phi$}{v1,v2}
\end{fmfgraph*}
\begin{fmfgraph*}(100,75)
\fmfleft{i1,i2}
\fmfright{o1,o2}
\fmf{plain, label=$\nu$}{v1,i2}
\fmf{plain, label=$\nu$}{v1,i1}
\fmf{plain, label=$\nu$}{v2,o2}
\fmf{plain, label=$\nu$}{v2,o1}
\fmf{dashes, label=$\phi$}{v1,v2}
\end{fmfgraph*}
\end{fmffile}
\vspace{1.0cm}
\begin{fmffile}{feynman2}
\begin{fmfgraph*}(100,75)
\fmfleft{i1,i2}
\fmfright{o1,o2}
\fmf{plain, label=$\nu$}{v1,i1}
\fmf{dashes, label=$\phi$}{o1,v1}
\fmf{dashes, label=$\phi$}{o2,v2}
\fmf{plain, label=$\nu$}{i2,v2}
\fmf{plain, label=$\nu$}{v1,v2}
\end{fmfgraph*}
\begin{fmfgraph*}(100,75)
\fmfleft{i1,i2}
\fmfright{o1,o2}
\fmf{plain, label=$\nu$}{v1,i2}
\fmf{plain, label=$\nu$}{v2,o2}
\fmf{dashes, label=$\phi$}{v1,i1}
\fmf{dashes, label=$\phi$}{v2,o1}
\fmf{plain, label=$\nu$}{v1,v2}
\end{fmfgraph*}
\end{fmffile}
\caption{Feynman diagrams for the binary processes allowed by the Lagrangian (\ref{eq:lagr}). Time goes from
left to right. From left to right: $\nu$-$\nu$ scattering (s and t channels), $\nu$-$\phi$ scattering,
$\nu\bar\nu$ annihilation to $\phi$'s. }
\label{fig:feyn}
\end{figure}

We consider neutrinos interacting with a light boson $\phi$ through simple scalar $h_{ij}$ and pseudoscalar $g_{ij}$ couplings,
as described by the following Lagrangian \cite{Chikashige:1980ui,Schechter:1981cv,Gelmini:1980re}:

\begin{equation}
{\mathcal L} =  h_{ij}\bar\nu_i\nu_j\phi+g_{ij} \bar \nu_i \gamma_5 \nu_j \phi + h.c. \,,
\label{eq:lagr}
\end{equation}

where the indices $i,\, j$ run over the neutrino mass eigenstates. This kind of interaction allows for the binary processes 
shown in Fig. \ref{fig:feyn}, i.e. $\nu + \bar \nu \leftrightarrow \phi+\phi$ (neutrino
annihilation to $\phi$'s),
$\nu + \phi  \leftrightarrow \nu + \phi$ (neutrino-$\phi$ scattering), $\nu + \nu \leftrightarrow \nu + \nu$ (neutrino-neutrino scattering mediated by a scalar boson exchange), as well as for neutrino decay $\nu \to \nu' + \phi$. 


Neutrino scalar and pseudoscalar couplings are constrained by laboratory
searches for neutrinoless double beta decay ($0\nu\beta\beta$), and by supernovae observations.
For example, in addition to the simplest $0\nu\beta\beta$ decay mode
\begin{equation}
(A,\,Z) \to (A,\,Z+2) + 2 e^{-} \, ,
\end{equation}
whose existence only requires the neutrino to be a Majorana particle \cite{Schechter:1981bd},
modes in which one or two additional $\phi$ bosons  are emitted
\begin{align}
&(A,\,Z) \to (A,\,Z+2) + 2 e^{-} +\phi\, , \\
&(A,\,Z) \to (A,\,Z+2) + 2 e^{-} +2\phi\, ,
\end{align}
are possible if neutrinos possess (pseudo)scalar couplings. 
$0\nu\beta\beta$ experiments
yield constraints on the effective $\phi$-neutrino coupling constant 
$\langle g_{ee} \rangle < (0.8 - 1.6) \times 10^{-5}$, depending
on the theoretical model \cite{Gando:2012pj,Albert:2014fya}.
The quantity $g_{ee}$ is the $e-e$ entry
of the coupling matrix in the weak base, related 
to the couplings $g_{ij}$ in the mass basis through the elements 
of the neutrino mixing matrix.

Neutrino decays $\nu \to \nu' + \phi$ can also be important in the high-density
supernova environment \cite{Choi:1989hi,Kachelriess:2000qc,Tomas:2001dh,Farzan:2002wx}. 
In the case of Majoron models,
limits on Majoron-neutrino couplings from observations of SN 1987A were derived in Ref. \cite{Kachelriess:2000qc}.
It has been shown there that $\phi$ emission would shorten too much the observed 
neutrino signal from SN 1987A if $3\times 10^{-7} \lesssim g \lesssim 2\times 10^{-5}$ 
(here $g$ denotes the largest element of the coupling matrix $g_{\alpha\beta}$ in 
the weak base), thereby excluding this region. Moreover, the observed $\bar \nu_e$ flux
from SN 1987A can also be used to further constraint $g_{11}\lesssim 10^{-4}$.
These limits, together with those from $0\nu\beta\beta$ decay experiments 
available at that time, were combined and translated into the mass basis in Ref. \cite{Tomas:2001dh}.

Scalar and pseudoscalar neutrino couplings can also be relevant in a cosmological context,
since collisional processes induced by the new interaction would affect the evolution
of perturbations in the cosmological neutrino fluid.
In general, the cross section for a binary process mediated by a massless boson
has the form $\sigma_\mathrm{bin} \sim g^4/s$ in the ultrarelativistic limit (apart from numerical factors)
with $g$ being the 
largest value of the Yukawa matrix (we do not distinguish between scalar and
pseudoscalar couplings in the following), and $\sqrt{s}$ is the center-of-mass energy.
Thus, in thermal equilibrium, the rate for a binary process is
\begin{equation}
\Gamma_\mathrm{bin} = \langle \sigma_\mathrm{bin} v \rangle n_\mathrm{eq} \propto g^4 T \, ,
\label{eq:rate_bin}
\end{equation}
since the equilibrium neutrino abundance $n_\mathrm{eq} \propto T^3$, and $s\sim T^2$. 

Interactions are of cosmological significance when the ratio 
$\Gamma/H$ between the interaction and Hubble expansion 
rates is of order unity or larger. The expansion rate scales as $H \sim T^2$ ($H\sim T^{3/2}$)
during the radiation- (matter-) dominated era.
The fact that the interaction rate (\ref{eq:rate_bin}) scales like $T$ has interesting phenomenological consequences:
since the ratio $\Gamma_\mathrm{bin}/H$ actually \emph{increases} with decreasing temperature, 
neutrinos, having decoupled in the early Universe, could recouple at late times
due to scattering/annihilation processes mediated by $\phi$.

Let us describe in more detail what happens once neutrino secret interactions become effective. 
We define the recoupling redshift of neutrinos $\zrec$,
through the condition $\Gamma_\mathrm{bin}(\zrec) = H(\zrec)$. At $z \lesssim \zrec$, the neutrino free-streaming
length rapidly drops well below the Hubble length due to scatterings. Hence, the neutrino contribution to the
cosmic shear becomes negligible. This effect should be observable in the CMB anisotropy spectrum, provided
recoupling happens close enough to recombination. Once recoupling becomes effective, one should expect
production of $\phi$'s through $\nu\bar\nu$ annihilation, and their subsequent thermalization through $\nu\phi$ scatterings\footnote{We 
neglect the population of $\phi$ bosons produced during reheating and the correspoding contribution to $\neff$,
since their energy density would be diluted by the entropy produced by standard model particles annihilations.}.
This allows us to describe the tightly coupled $\nu$'s and $\phi$'s as a single fluid. Note that if 
late $\phi$ production, as we assume, happens when both neutrinos and $\phi$'s are in the ultrarelativistic regime,
the total energy stored in relativistic species does not change, so $\neff$ remains constant.
This holds as long as chemical equilibrium is
mantained. Once the temperature of the fluid falls below the neutrino mass,
annihilation processes start to deplete the neutrino abundance, and the energy stored in the neutrino rest mass
would end up increasing $\neff$ \cite{Hannestad:2004qu,Archidiacono:2014nda}. We neglect this effect in what follows, assuming that neutrino masses are small enough to become cosmologically 
significant well after recoupling. Finally, we also neglect the possibility of neutrino decay, implicitly assuming 
that the off-diagonal couplings are small.

\section{The Bolzmann equation} \label{sec:boltz}

\subsection{Formalism}

The phase-space evolution of the components of the primordial plasma
is described by the Boltzmann equation,
that can be generically written as:
\begin{equation}
\frac {D f}{D \tau}= \hat C[f] \, ,
\label{eq:boltz}
\end{equation}
where $f$ is the phase-space distribution function (DF), and $\hat C[f]$ is the collision operator, describing 
interactions between particles, and $\tau$ is a generic time variable, that we here take to be conformal time.
In fact, the cosmological evolution
is described by a set of Boltzmann equations, one for each component
of the cosmological fluid, coupled between them by gravity [hidden in the LHS
of Eq. (\ref{eq:boltz})], and possibly by the collision term (as in the case of baryons and photons, 
coupled by Thomson scattering). This system of equations must be complemented 
by the Einstein equations describing the evolution of metric variables.

In a perturbed Friedmann-Robertson-Walker Universe, the DF can be conveniently written as the sum
of an unperturbed part $f_0$ (that does not depend on position nor
on the direction of momentum, due to the homogeneity and isotropy of the background spacetime) 
plus a small perturbation $\delta f \equiv f_0\Psi$:
\begin{equation}
f(\Vec{x}, \Vec{q}, \tau)= f_0(q) [1+ \Psi(\Vec{x}, \Vec{q}, \tau)] \, ,
\end{equation}
where we are using comoving coordinates $x^i$ and momenta $q_i \equiv q n_i$ ($n_i$ being a unit vector) as 
the spatial and momentum coordinates respectively.

In the synchronous gauge, the perturbed metric is written in the form 
\begin{equation}
ds^2 = a^2(\tau) \left[ -d \tau^2 + \left( \delta_{ij} + h_{ij} \right) dx^i dx^j \right] ,
\end{equation}
where $a(\tau)$ is the cosmic scale factor and $h_{ij}$ are the metric perturbations. With
this choice of the gauge, the perturbed Boltzmann equation in $k$-space takes the form
(for the sake of simplicity, we use the same symbol for $\Psi$ and its Fourier transform):
\begin{equation}
\frac{\partial \Psi}{\partial \tau} +  i k\mu \frac{q}{\epsilon}  \Psi 
+ \frac{d \ln f_0}{d \ln q} \left[\dot{\eta} - \frac{\dot{h}+6 \dot{\eta}}{2} \mu^2  \right]  = 
\frac{1}{f_0} \hat C[f] \, ,
\label{eq:boltzk}
\end{equation}
where $\epsilon = \sqrt{q^2+a^2m^2}$ ($m$ being the mass of the particle), $\mu \equiv \hat k \cdot \hat n$ is the angle 
between the perturbation wave number and the particle momentum, and $h$ and $\eta$ 
are the scalar components of the metric perturbation $h_{ij}$ in $k$-space:
\begin{equation}
h_{ij} (\vec x, \,\tau) = \int d^3 k\,e^{i \vec k \cdot \vec x} \Big[\hat k_i \hat k_j h(\vec k,\,\tau) + 
\big(\hat k_i \hat k_j -\frac{1}{3} \delta_{ij}\big)\, 6\eta(\hat k,\,\tau)\Big]
\end{equation}
For the moment, let us consider a collisionless fluid, \textit{i.e.}, $\hat C[f]=0$. In
the case of massless particles, $\epsilon = q$ and the Boltzmann equation (\ref{eq:boltzk})
can be further simplified by integrating out the momentum dependence of the DF. Defining
\begin{equation}
F(\vec k,\,\hat n,\,\tau) \equiv \frac{\int q^3 f_0(q) \Psi(\vec{k},\,\vec q,\,\tau)\,dq}{\int q^3 f_0(q)\,dq} \, ,
\end{equation}
and further expanding the angular dependence of $F$ in a series of Legendre polynomials:
\begin{equation}
F(\vec k,\,\hat n,\,\tau) = \sum_{\ell=0}^{\infty} (-i)^\ell (2\ell+1) F_\ell(\vec k, \tau) P_\ell (\mu) \,
\end{equation}
the following Boltzmann hierarchy is obtained:
\begin{subequations}
\begin{align}
&\dot{\delta} = -\frac{4}{3} \theta - \frac{2}{3} \dot{h} \, ,\\
&\dot{\theta} = k^2 \left(\frac{1}{4} \delta - \Pi \right) \, ,\\ 
&\dot{\Pi} = \frac{4}{15} \theta- \frac{3}{10} k F_{3} + \frac{2}{15} \dot{h} + \frac{4}{5}\dot{\eta} \, ,\\
&\dot{F_\ell} = \frac{k}{2\ell+1} \Big[ \ell F_{\ell-1} - (\ell+1) F_{\ell+1} \Big] \quad (\ell\ge 3) \, ,
\end{align}
\label{eq:boltz_hierarchy}
\end{subequations}
where $\delta \equiv F_0$, $\theta \equiv (3/4) k F_1$ and $\Pi \equiv F_2/2$
(our $\Pi$ corresponds to $\sigma$ in the notation of Ref. \cite{Ma:1995ey}).

\subsection{Boltzmann equation for interacting neutrinos}
\label{subsec:int_boltz}

In order to study the behaviour of non-standard interacting neutrinos,
we need to specify the form of the collision term on the RHS side of 
the Boltzmann equation. In principle, we should compute the collision integrals
given a specific form of the interaction lagrangian. Detailed calculations of the 
collision term for neutrino-neutrino interactions
mediated by a scalar particle have been presented, under some assumptions, 
in Ref. \cite{Oldengott:2014qra}.
In that work, the collision integrals are simplified by reducing their dimensionality,
but yet they cannot be computed analytically, leaving the Boltzmann equation 
in its integro-differential form. Moreover, the presence of the collision terms
prevents from integrating out the momentum dependence in the Boltzmann equation
 also in the massless case, increasing the numerical complexity of the problem.

In this work, instead, we will pursue a simpler approach and use the relaxation time
approximation as done in  \cite{Hannestad:2000gt}. This approximation amounts
in writing the collision integral as a damping term, proportional to the inverse
mean free time between collisions $\tauc$, i.e.
\begin{equation}
\hat C[f] \propto - \frac{1}{\tauc} \delta f \, .
\label{eq:reltime}
\end{equation}
where $\tauc^{-1} = a \Gamma =  a n \langle \sigma v \rangle$.
Here $\Gamma = n \langle \sigma v \rangle $ is the interaction rate in the comoving frame, and the scale factor
comes from the fact that we choose conformal time as the time coordinate.

Following the same steps that led from Eq. (\ref{eq:boltzk}) to Eqs. (\ref{eq:boltz_hierarchy}),
it is straighforward to show that a collision term of the form (\ref{eq:reltime})
would result in additional terms proportional to $- a \Gamma F_\ell$ on the RHS of the
equation for $\dot F_\ell$. Notice, however, that conservation of the number of particles
(as in a $2\leftrightarrow2$ scattering process) and conservation of momentum imply
respectively
\begin{align}
&\int q^3 \hat C[f] \,d\Omega\, dq = 0 \, , \\
&\int \mu q^3 \hat C[f] \,d\Omega\,\,dq  = 0 \, ,
\end{align}
\emph{i.e.}, no collision terms should appear in the monopole and dipole evolution equations.
For this reason we set to zero the collision terms for $\ell =0,\,1$ , so that
the Boltzman hierarchy for interacting neutrinos reads [we choose the proportionality
constant in Eq. (\ref{eq:reltime}) to be unity]
\begin{subequations}
\begin{align}
&\dot{\delta} = -\frac{4}{3} \theta - \frac{2}{3} \dot{h} \, ,\\
&\dot{\theta} = k^2 \left(\frac{1}{4} \delta - \Pi \right) \, ,\\ 
&\dot{\Pi} = \frac{4}{15} \theta- \frac{3}{10} k F_{3} + \frac{2}{15} \dot{h} + \frac{4}{5}\dot{\eta} - a \Gamma \Pi\, ,\\
&\dot{F_\ell} = \frac{k}{2\ell+1} \Big[ \ell F_{\ell-1} - (\ell+1) F_{\ell+1} \Big]  -a \Gamma F_\ell \quad (\ell\ge 3)\, .
\end{align}
\label{eq:boltz_hierarchy_int}
\end{subequations}
We have noted above that, once secret interactions become cosmologically significant, neutrino-neutrino annihilations
start to produce $\phi$ bosons that are rapidly brought into thermal equilibrium with neutrinos themselves by scatterings.
This allows to treat the neutrino-$\phi$ system as a single, tightly coupled self-interacting fluid, whose perturbations still evolve
according to Eqs. (\ref{eq:boltz_hierarchy_int}). The conservation of particle number and momentum 
still holds, at the level of the coupled fluid, justifying the vanishing $\ell =0,\,1$ collision terms also in this regime.

\section{Method}
\label{sec:method}

In the following, we parameterize the strength of the non-standard interaction 
by generically writing the coefficient of the collisional damping terms appearing 
in the RHS of Eqs. (\ref{eq:boltz_hierarchy_int}) for $\ell\ge 2$  as
\begin{equation}
\Gamma_\mathrm{bin} = \gnn^4 T_\nu \, .
\label{eq:cambpar_Gamma}
\end{equation}
This has the right energy dependence for an interaction mediated by a light (pseudo)scalar boson in
the ultrarelativistic limit. With such a choice, the parameter $\gnn$ is roughly proportional, through a dimensionless factor,
to the largest Yukawa coupling appearing in the Lagrangian (\ref{eq:lagr}). The exact relation between $\gnn$ 
and the $g_{ij}$'s depends on the details of the underlying particle physics model. 
Nevertheless, in the following, we shall loosely refer to $\gnn$ as the coupling constant for non-standard
neutrino interactions.

We use a modified version of the publicly available Boltzmann code \texttt{camb} \cite{camb} in order 
to solve the perturbation evolution and compute the CMB power spectra for 
given values of the cosmological parameters.
In particular, we have modified the evolution equations for interacting 
massless neutrinos described in Sec. \ref{subsec:int_boltz} in order to 
include a collision term with the form (\ref{eq:cambpar_Gamma}).

\subsection{Cosmological data analysis}

The baseline model considered in this paper is described by the following parameter set:
\begin{equation}
\label{parameter}
  \{\omega_b,\,\omega_c, \,\theta,\, \taureio,\, n_s,\, \log[10^{10}A_{s}],\, \gnn^4 \}~,
\end{equation}
where $\omega_b$ and $\omega_c$ are the physical baryon and cold dark matter
density, respectively, $\theta$ is the angle subtended by the sound horizon
at the time of recombination, $\taureio$ is the optical depth to reionization, $n_s$ and $A_s$
are the spectral index and amplitude of the primordial spectrum of scalar
fluctuations (both evaluated at the pivot wavenumber $k_0 = 0.05~\Mpc^{-1}$),
and $\gnn^4$ parameterizes the strength of non-standard neutrino interactions, 
as per Eq. ({\ref{eq:cambpar_Gamma}). These are the parameters that are varied
in the MC run and that implicitly take flat priors. We assume flat geometry, massless neutrinos and
adiabatic initial conditions. In the baseline model, we also fix $\neff = 3.046$  and ignore tensor modes by setting the 
tensor-to-scalar ratio $r=0$. We refer to this model as $\Lambda$CDM+$\gnn$. 
When $\gnn$ is fixed to zero, it reduces to the standard
$\Lambda$CDM model with weakly-interacting neutrinos. We also consider extensions
to the baseline model, by allowing $\neff$ or $r$ to vary, referred to as $\Lambda$CDM+$\gnn$+$\neff$ 
and $\Lambda$CDM+$\gnn$+$r$, respectively.
In the following we shall also presents results for derived parameters of interest, like the present Hubble parameter
$H_0$, the tensor-to-scalar ratio $r_{0.002}$ evaluated at $0.002\,\Mpc^{-1}$ and the
neutrino recoupling redshift $\zrec$. In Tab. \ref{tab:priors} we summarize 
our choice of parameterization. In addition to the parameters listed in the table, we also vary
a number of ``nuisance'' parameters describing residual foregrounds and instrumental characteristics.

\begin{table}
\caption{Cosmological parameter used in the analysis. The upper part of the table lists the base parameters, \emph{i.e.}, those with
uniform priors that are varied in the Monte Carlo runs. In the baseline model we only consider the first seven of those listed here; $\neff$ and $r$ are varied in extensions of this model. The lower part lists derived parameters of interest, for which we also compute credible intervals.
For each parameter, we quote the initial prior range (for base parameters only).}
\begin{center}
\begin{tabular}{l l c} 
Parameter				&	\quad Definition			&Prior range  \\
\hline \hline
$\Omega_\mathrm{b} h^2$ 	&\quad  Present baryon density		& [0.005, 0.1] \\
$\Omega_\mathrm{c} h^2$ 	&\quad  Present dark matter density	& [0.005, 0.1] \\
100 $\theta$					& \quad Angular size of the sound horizon at recombination \quad		& [0.5, 10] \\
$\taureio$			& \quad Optical depth to recombination	& [0.01, 0.8] \\
$n_s$					& \quad Spectral index of scalar perturbations 			&  [0.5, 1.5] \\
$\ln(10^{10} A_s)$					& \quad Log amplitude of scalar perturbations at $k_0=0.05$ Mpc$^{-1}$ \qquad & [2.7, 4.0]  \\
$10^{27} \gnn^4$		&\quad Strength of non-standard neutrino interactions\footnote{See definition in Eq. (\ref{eq:cambpar_Gamma}).} \qquad & [0, 2]\\[0.1cm]
$\neff$ & \quad Effective number of neutrino families\footnote{Fixed to $\neff=3.046$ in the baseline model. We use $\neff$ to refer
to the value before scalars are produced by $\nu\bar\nu$ annihilations - see discussion in the text.} \quad & [0, 5] \\
$r$ & \quad Tensor-to-scalar ratio at $k_0=0.05$ Mpc$^{-1}$ \footnote{Fixed to $r=0$ in the baseline model. We assume
the inflation consistency relation $n_t = - r/8$ for the spectral index $n_t$ of the primordial spectrum of tensor fluctuations.} \qquad & [0, 2]\\
\hline
& & \\[-0.2cm]
$H_0$ &\quad Hubble constant at $a(t)=1$ \quad & - \\
$r_{0.002}$ & \quad Tensor-to-scalar ratio at $k_0=0.002$ Mpc$^{-1}$ \quad & - \\[0.1cm]
$\zrec$ & \quad Redshift of neutrino recoupling \quad & - \\[0.1cm]
\hline
\end{tabular}
\end{center}
\label{tab:priors}
\end{table}%

In order to derive constraints for the parameters of the model,
we perform a Markov Chain Monte Carlo (MCMC)
analysis through the publicly available \texttt{CosmoMC} code \cite{Lewis:2002ah},
interfaced with our modified version of camb. \texttt{CosmoMC} uses
a modified version of the Metropolis-Hastings algorithm \cite{Lewis:2013hha}
to sample the full posterior distribution of the parameters given the data 
(described in the following section).
Lower-dimensional posterior distributions are obtained by marginalizing over
unwanted parameters. When quoting credible intervals, we apply the following 
rule: if both edges of the 95\% credible interval are distinct from the prior edges,
we quote constraints in the form mean $\pm$ 68\% uncertainty; on the contrary
we quote the 95\% upper or lower limit.

\subsection{Cosmological data}

We consider the data on CMB temperature anisotropies released
by the Planck satellite in 2013 \cite{Ade:2013sjv,Ade:2013kta},
supplemented by the 9-year polarization data from WMAP \cite{Bennett:2012zja},
as well as additional temperature data from high-resolution CMB experiments,
namely the Atacama Cosmology Telescope (ACT) \cite{Das:2013zf} and 
the South Pole Telescope (SPT) \cite{Reichardt:2011yv}. The purpose of 
considering the ACT and SPT data is mainly to improve constraints on the unresolved foregrounds. 

The likelihood functions associated to these datasets are evaluated and combined
using the likelihood code
distributed by the Planck collaboration, described in Ref. \cite{Ade:2013kta}, and publicly
available at Planck Legacy Archive\footnote{\texttt{http://pla.esac.esa.int/pla/}}.
This likelihood uses Planck TT data up to a maximum multipole number of $\ell_{\rm max}=2500$, 
and WMAP 9-year polarization data  (WP) up to $\ell=23$ \cite{Bennett:2012zja}, as well as
ACT data in the range $1000<\ell<9440$ \cite{Das:2013zf} and SPT data in the range $2000<\ell<10000$ \cite{Reichardt:2011yv}.

We also use the likelihood recently released by the joint Planck and BICEP2/Keck effort
\cite{Ade:2015tva}. This likelihood is based on CMB polarization observations from 
the BICEP2 field and uses corresponding data from Planck 2015 217 and 353 GHz channels to 
account for contamination from polarized Galactic dust. It is limited to the multipole range $20 < \ell < 200$.

\section{Results}

\label{sec:res}

\subsection{Impact on the evolution of pertubations}

Once neutrinos become interacting again, their anisotropic stress (as well as all the higher-order moments of 
the distribution function) is suppressed and eventually vanishes once the
tightly coupled regime ($\Gamma_{\nu\nu}/H \to \infty$) is reached. At the same time,
the damping of density perturbation caused by neutrino free streaming is no more effective and the neutrino fluid
undergoes acoustic oscillations. In order to illustrate this, in Fig. \ref{fig:perturb-evolution} we 
show the evolution of density ($\delta_\nu$) and shear ($\Pi_\nu$) perturbations for three
different wave numbers ($k = 5 \times 10^{-3},\, 5 \times 10^{-2}, \,5 \times 10^{-1}\,\Mpc^{-1}$), in the case of non-interacting neutrinos (\emph{i.e.}, $\gnn=0$), and for two finite values of
the secret interaction strength, namely $\gnn =  1.2 \times 10^{-7},\,2 \times 10^{-7}$.
The latter values correspond to a redshift of neutrino decoupling $\zrec\simeq 1300$ and $1.8\times 10^4$, respectively (in this subsection we take the following
benchmark for the cosmological parameters: $\omega_b = 0.0227$, $\omega_c = 0.124$, $h=0.68$, $\taureio = 0.09$, $n_s =0.96$, $A_s = 2.1\times 10^{=9}$, $\neff = 3.046$, $r=0$). 
The upper panels of Fig. \ref{fig:perturb-evolution} show perturbation evolution for the largest scale, $k=5 \times 10^{-3}\,\Mpc^{-1}$, entering the horizon around the time of hydrogen recombination at $z\simeq 1100$. For the larger value of the coupling constant, when the mode enters the horizon neutrinos are already completely recoupled, and shear oscillations are overdamped. For $\gnn =  2 \times 10^{-7}$, on the other hand, neutrinos are only partially coupled at the time of horizon crossing and this results in a significant but not complete suppression of shear perturbations with respect to the non-interacting case. The evolution of density perturbations mirrors that of the shear: when the dissipation normally associated to neutrino free-streaming is absent, undamped acoustic oscillations set on in the fluid, so that density perturbations are actually boosted by increasing $\gnn$. In the middle row, we show results for $k=5 \times 10^{-2}\,\Mpc^{-1}$. This
scale crosses the horizon at $z\simeq 2 \times 10^4$. In this case, the shear is again suppressed as soon as the mode enters the horizon for $\gnn =  2\times 10^{-7}$, since neutrinos are recoupling roughly at the same time, 
while for $\gnn =  1.2 \times 10^{-7}$ the effect of interactions only kicks off after 
perturbations have been damped by the expansion. Finally, for the smallest scale under consideration
($k =0.5 \,\Mpc^{-1}$), shown in the bottom row and entering the horizon at $z\simeq 2\times 10^5$, neutrino recoupling
happens in both cases when the perturbation is well inside the horizon, making the evolution very similar
to the non-interacting case.

\begin{figure}
\includegraphics[width=1.05\textwidth,keepaspectratio]{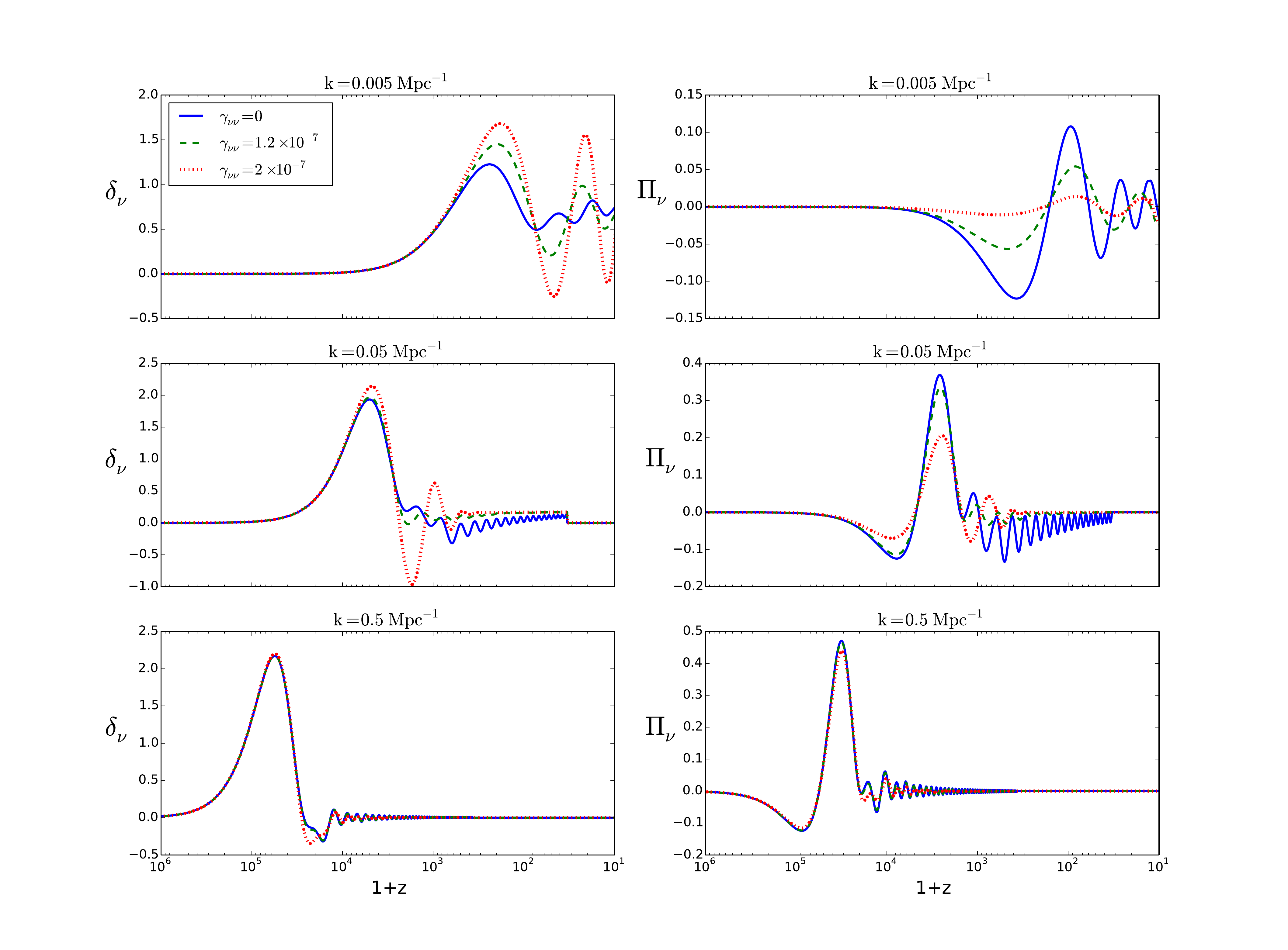}
\caption{Evolution of neutrino density (left column) and shear (right column) perturbations for three different modes $k = 0.005,\, 0.05,\, 0.5\, \Mpc^{-1}$. We compare the evolution for standard, weakly-interacting neutrinos (blue solid curves) with the case of non-standard neutrino
interactions with $\gnn =1.2 \times 10^{-7}$ and $\gnn = 2 \times 10^{-7}$ (green dashed and red dotted curves respectively). Non-standard interactions suppress shear perturbations while boosting density perturbations after the time of neutrino recoupling (see text for discussion).}
\label{fig:perturb-evolution}
\end{figure}

In Fig. \ref{fig:psTT}, we show how secret neutrino interactions affect the CMB power spectra. 
We have computed the power spectra for the same models shown in Fig. \ref{fig:perturb-evolution},
i.e. $\gnn = 0,\, 1.2 \times 10^{-7},\,2 \times 10^{-7}$. The effect of neutrino interactions on the temperature power spectrum
shown in the upper panel of Fig. \ref{fig:psTT} is
to boost the overall amplitude of the spectrum as $\gnn$ is increased.  The increased power 
is caused by the increased density fluctuations due to the absence of neutrino free streaming.
A similar behaviour is observed in the TE power spectrum, as shown in the lower panel
of the same figure.

\begin{figure}[t]
\begin{center}
\includegraphics[width=0.48\linewidth,keepaspectratio]{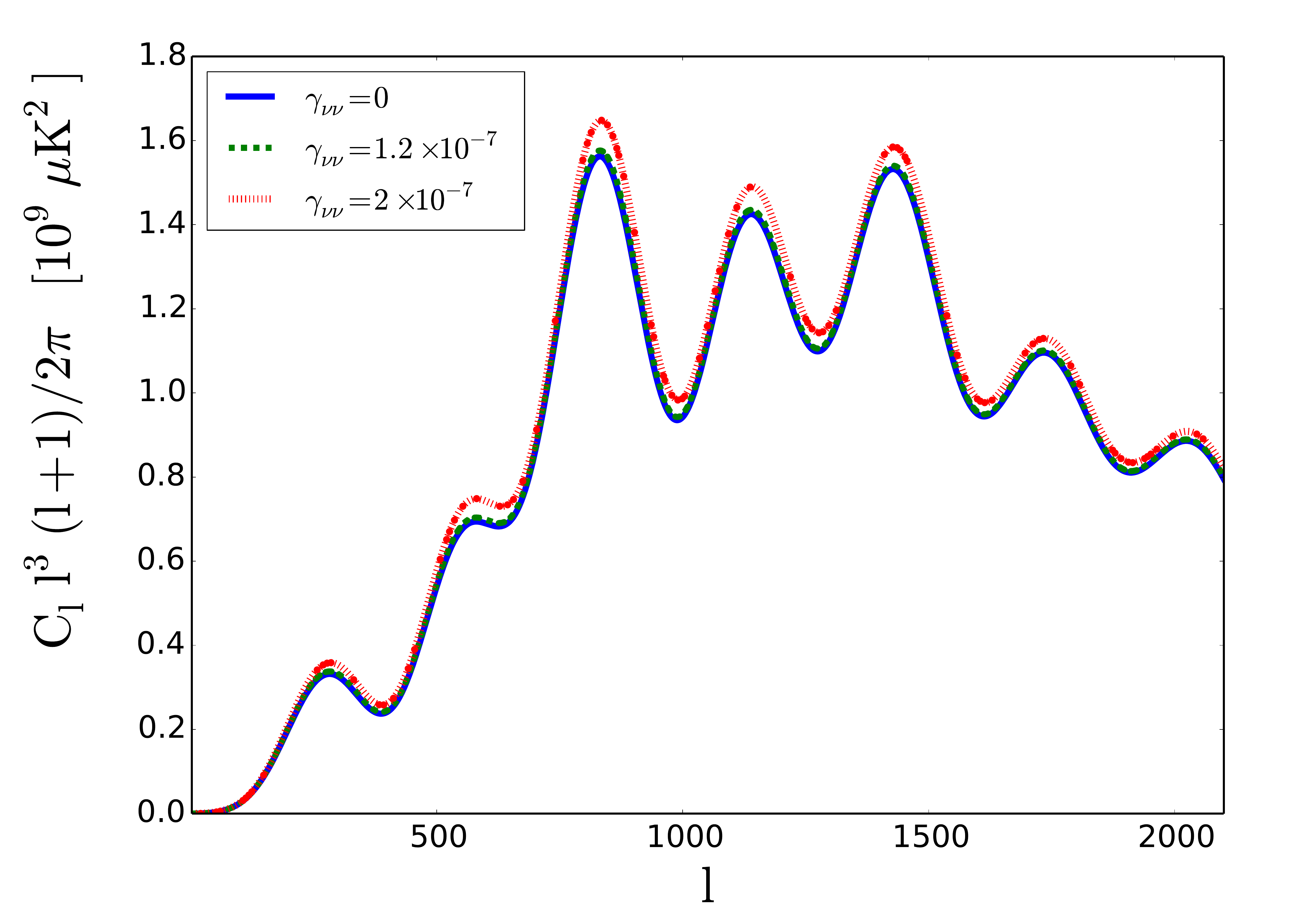}
\includegraphics[width=0.48\linewidth,keepaspectratio]{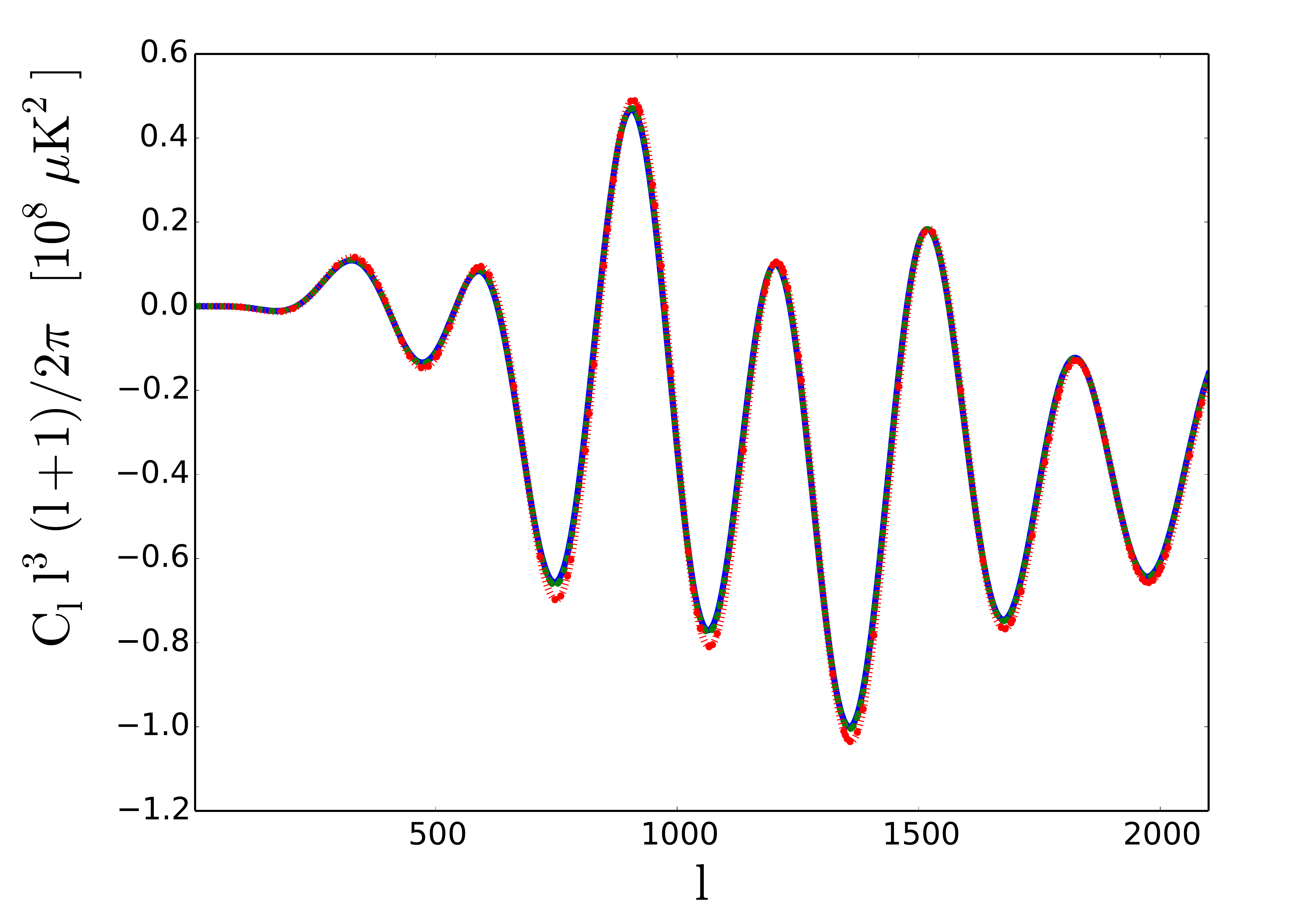}
\caption{CMB temperature (left panel) and temperature-$E$ polarization (right panel) power spectra for neutrinos with non-standard interactions.
As in Fig. \ref{fig:perturb-evolution} we show results for $\gnn = \{0,\, 1.2 \times 10^{-7},\,2 \times 10^{-7}\}$.}
\label{fig:psTT}
\end{center}
\end{figure}

\subsection{Constraints on cosmological parameters}

We are now ready to present the constraints that CMB data provide on the non-standard coupling constant $\gamma_{\nu \nu}$. In the following,
we will quote 68\% CL uncertainties, unless we are dealing with upper limits, in which case we quote 95\% credible intervals. The results
shown in the following are summarized in Tabs. \ref{tab:pars1} and \ref{tab:pars2}.

Let us start by considering the simplest extension of the standard cosmological model, labelled $\mathrm{\Lambda CDM + \gamma_{\nu \nu}}$. Considering only Planck temperature and WMAP polarization data (Planck+WP), we obtain $\gnn^4 < 0.85 \times 10^{-27}$, or equivalently $\gnn < 1.71\times10^{-7}$, which implies a neutrino-neutrino recoupling at $\zrec \lesssim 8800$ (here and for the rest of the section, we will fix the other parameters to their best estimates when translating limits on $\gnn$ to limits on $\zrec$). Adding the ACT and SPT datasets (``highL'') shifts the distribution to larger values of the coupling constant, yielding $\gnn^4 < 0.96\times 10^{-27}$ ($\gnn < 1.76\times10^{-7}$ and $\zrec \lesssim 10^4$). In the upper panel of Fig. \ref{fig:gnn_post} we show the posterior distributions for $\gnn^4$ in the $\Lambda$CDM+$\gnn$ model, for the Planck+WP and Planck+WP+highL datasets. Both posteriors are quite asymmetric and 
have a peak at non-zero values of the coupling constant, respectively at $\gnn^4 = 0.24\times10^{-27}$ and $0.36\times10^{-27}$, corresponding to $\zrec\simeq2800$ and $1700$. Interestingly enough, there is a weak (at the $\sim 1\sigma$ level) preference for non-zero values of $\gnn^4$: at 68\% CL, we find $\gnn^4 = (0.369^{+0.082}_{-0.367})\times10^{-27}$ (Planck+WP) and $\gnn^4 =(0.444^{+0.169}_{-0.356})\times10^{-27}$ (Planck+WP+HL).
The $68\%$ lower limit in the latter case corresponds to $\zrec \gtrsim 300$.

We have also constrained the number of relativistic species in conjuction with $\gnn$.
In the framework of this $\Lambda$CDM+$\gamma_{\nu \nu} + \neff$ model, we find a 95\% credible interval 
$\gnn^4 <0.75\times10^{-27}$, or $\gnn <1.65\times10^{-7}$ from Planck+WP. This value provides a neutrino-neutrino 
recoupling at $\zrec \lesssim 7400$.   
Also in this case, the addition of the ACT and SPT datasets  weakens the constraints on the
coupling constant, yielding $\gnn^4 <0.93\times10^{-27}$ ($\gnn <1.75\times10^{-7}$ and $\zrec \lesssim 9800$) at $95\%$ CL.
For what concerns the effective number of relativistic species, we find $\neff=3.44^{+0.37}_{-0.41}$ (Planck+WP) and $\neff=3.27^{+0.33}_{-	0.38}$
(Planck+WP+HL). This is very much consistent with the corresponding values found by the Planck collaboration
in the $\neff$ extension of the $\Lambda$CDM model for the same datasets \cite{Ade:2013zuv}.
The posterior distributions for $\gnn$ in the $\Lambda$CDM+$\gnn + \neff$ model, for the datasets under consideration
are presented in the right pannel of Fig.\ref{fig:gnn_post}. The maximum probability is obtained for $\gnn^4 = 0.18\times 10^{-27}$ ($\zrec  \simeq 1000$) and
$\gnn^4 = 0.27\times 10^{-27}$ ($\zrec \simeq 2000$) for Planck+WP and Planck+WP+HL, respectively. 
The presence of additional relativistic degrees of freedom reduces the preference for 
non-zero values of the coupling constant: in the $\Lambda$CDM+$\gnn + \neff$ model, $\gnn^4$ is consistent
with zero at below the $\sim 1\sigma$ level for Planck+WP, while the 68\% interval for Planck+WP+HL is $\gnn^4 = 0.403^{+0.115}_{-0.363}$, 
thus shifted to lower values with respect to the corresponding interval in the $\Lambda$CDM+$\gnn$ model.
In Fig. \ref{fig:gnn_post2D}, we show the most significant
correlations between $\gnn^4$ and other parameters, namely $\Omega_c h^2$, $\theta$, $10^9 A_s e^{-2\tau}$
and $\neff$. The correlations with the angle $\theta$ subtended by the sound horizon at recombination and 
with the amplitude $10^9 A_s e^{-2\tau}$ are particularly evident.
We argue that
the pattern leading to these correlations is the following: the overall amplitude of the spectrum increases for larger
values of $\gnn^4$, while the position of peaks and dips remains unchanged. This can be directly compensated 
by a lower value of $10^9 A_s e^{-2\tau}$. Alternatively, increasing $\Omega_c h^2$ (or decreasing $\neff$ if the model allows), 
lowers the height of the first few peaks but shifts their position to lower multipoles;
increasing $\theta$ moves the peaks back to their original position. To support our reasoning,
we show in the left panel of Fig. \ref{fig:post3D} a scatter plot of samples from Planck+WP chains in the $\Omega_c h^2-\theta$ plane,
color-coded by $\gnn^4$, compared with the 68\% and 95\% confidence region for the $\Lambda$CDM model.
It is clear from this plot that models with $\gnn^4 > 0$ can be made in agreement with the data 
by increasing both $\Omega_c h^2$ and $\theta$. The right panel of the same figure shows the same information
in terms of $H_0$ instead than $\theta$.

\begin{table}[t!]
\begin{small}
\begin{center}
\begin{tabular}{c|c|c|c|c}
\hline\hline
 & \multicolumn{4}{c}{}\\
 & \multicolumn{2}{c}{$\Lambda$CDM+$\gnn$} & \multicolumn{2}{c}{$\Lambda$CDM+$\gnn+\neff$} \\[0.1cm]
   \cline{2-5}
 &                    &                               &                         &                                 \\[0.1cm]
 & Planck+WP          & Planck+WP                     & Planck+WP               &  Planck+WP                      \\
 &                    & +highL                     &                         &  +highL                      \\
Parameter             &                               &                         &                                 \\
\hline 
 &                    &                               &                         &                                 \\
$\Omega_{\textrm{b}}h^2$& $0.02214\pm0.00029$ & $ 0.02220 \pm 0.00029 $ &$0.02244\pm0.00041$ & $ 0.02237\pm0.00040 $    \\[0.1cm]
$\Omega_{\textrm{c}}h^2$&$0.1218\pm0.0029$   &$ 0.1220 \pm 0.0029$ &$0.1263\pm 0.0054$   &$ 0.1247 \pm0.0048$    \\[0.1cm]
100 $\theta$                &$1.04195^{+0.00074}_{-0.00085}$   &$ 1.04210^{+0.00076}_{-0.00083}$ &$1.04145^{+0.00084}_{-0.00097}$   &$ 1.04181^{+0.00085}_{-0.00103} $    \\[0.1cm]
$\taureio$                  &$0.091^{+0.013}_{-0.014}$   &$ 0.093^{+0.013}_{-0.014}$ &$0.096^{+0.014}_{-0.016}$   &$ 0.095^{+0.013}_{-0.016} $    \\[0.1cm]
$n_s$                   &$ 0.9641 \pm 0.0074$  &$ 0.9629 \pm 0.0072$ &$0.979\pm0.016$   &$ 0.971 \pm 0.015$    \\[0.1cm]
$\log[10^{10} A_s]$     &$3.079^{+0.025}_{-0.026} $  &$ 3.080^{+0.025}_{-0.027} $ &$3.102^{+0.033}_{-0.037}$     &$ 3.092^{+0.033}_{-0.036} $    \\[0.1cm]
$10^{27}\gnn^4$       &$<0.85$&$ <0.96 $ &$<0.75$& $ <0.93$    \\[0.1cm]
$\neff$                 &          3.046         &  $3.046$  &$3.44^{+0.37}_{-0.41}$     &  $3.27^{+0.33}_{-0.38}$     \\[0.1cm]
\hline
$H_0$ [km/sec/Mpc] & $67.4 \pm 1.2$ &$67.4\pm 1.2$  & $70.4^{+3.0}_{-3.4}$ & $69.1^{+2.7}_{-3.2}$ \\[0.1cm] 
$10^{7}\gnn$       &$<1.71$&$ <1.76 $ &$<1.65$& $ <1.75$    \\[0.1cm]
\hline\hline
\end{tabular}
 \caption{Constraints on cosmological parameters for the $\Lambda$CDM+$\gnn$ and $\Lambda$CDM+$\gnn$+$\neff$ models from the analysis of the Planck+WP and Planck+WP+highL datasets. We quote 68\% C.L., except for upper bounds, which are $95\%$~C.L. }
 \label{tab:pars1}
 \end{center}
 \end{small}
 \end{table}
\begin{figure}
\begin{center}
\includegraphics[scale=0.52]{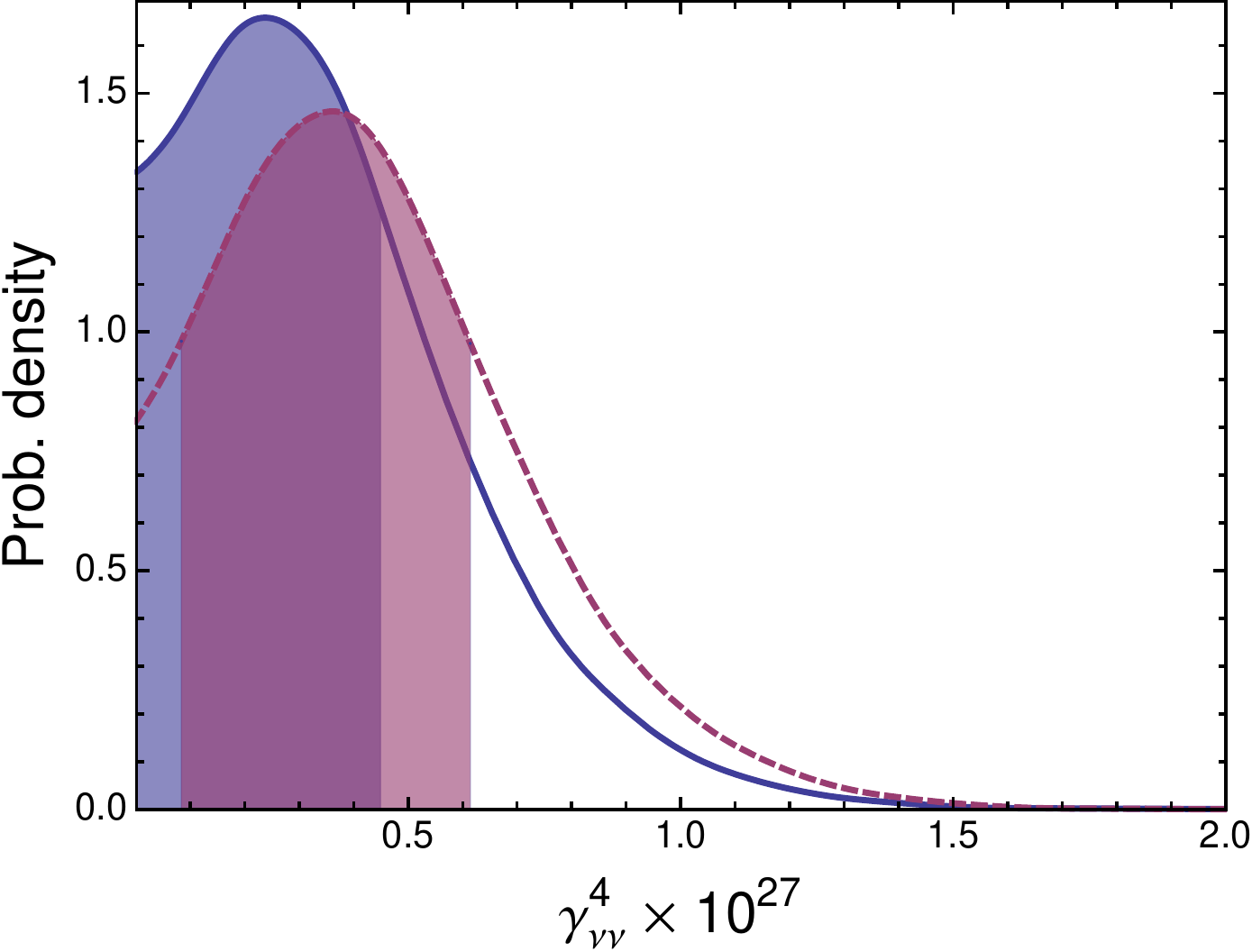}
\includegraphics[scale=0.52]{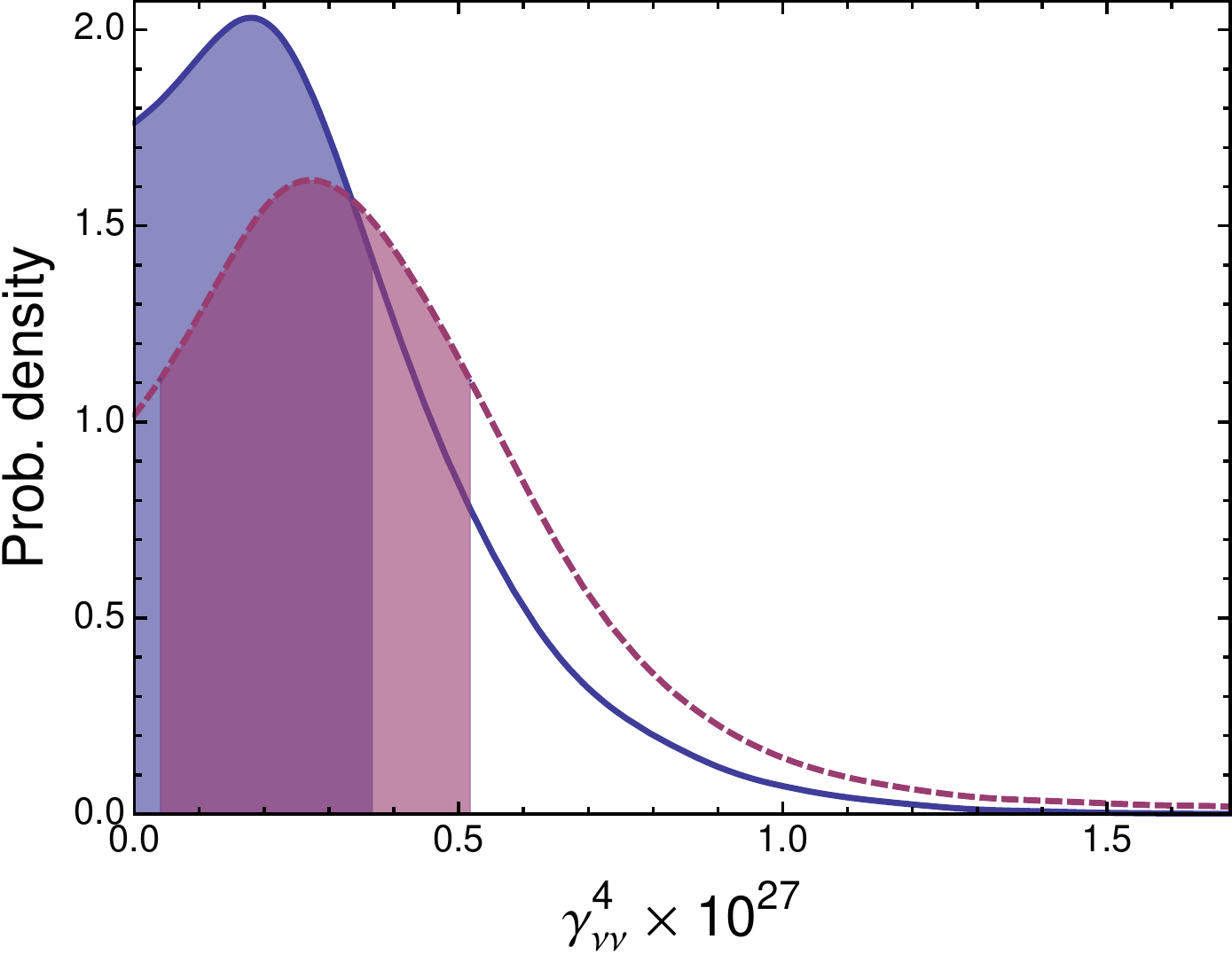}
\caption{One-dimensional posterior distribution for $\gnn^4$ obtained from the Planck+WP (blue solid) and Planck+WP+highL (red dashed)
datasets. The shaded regions denote the 68\% credible interval. Left panel: $\Lambda$CDM + $\gnn$. Right panel: $\Lambda$CDM+$\gnn$+$\neff$.}
\label{fig:gnn_post}
\end{center}
\end{figure}

\begin{figure*}
\begin{center}
\includegraphics[width=1.\textwidth,keepaspectratio]{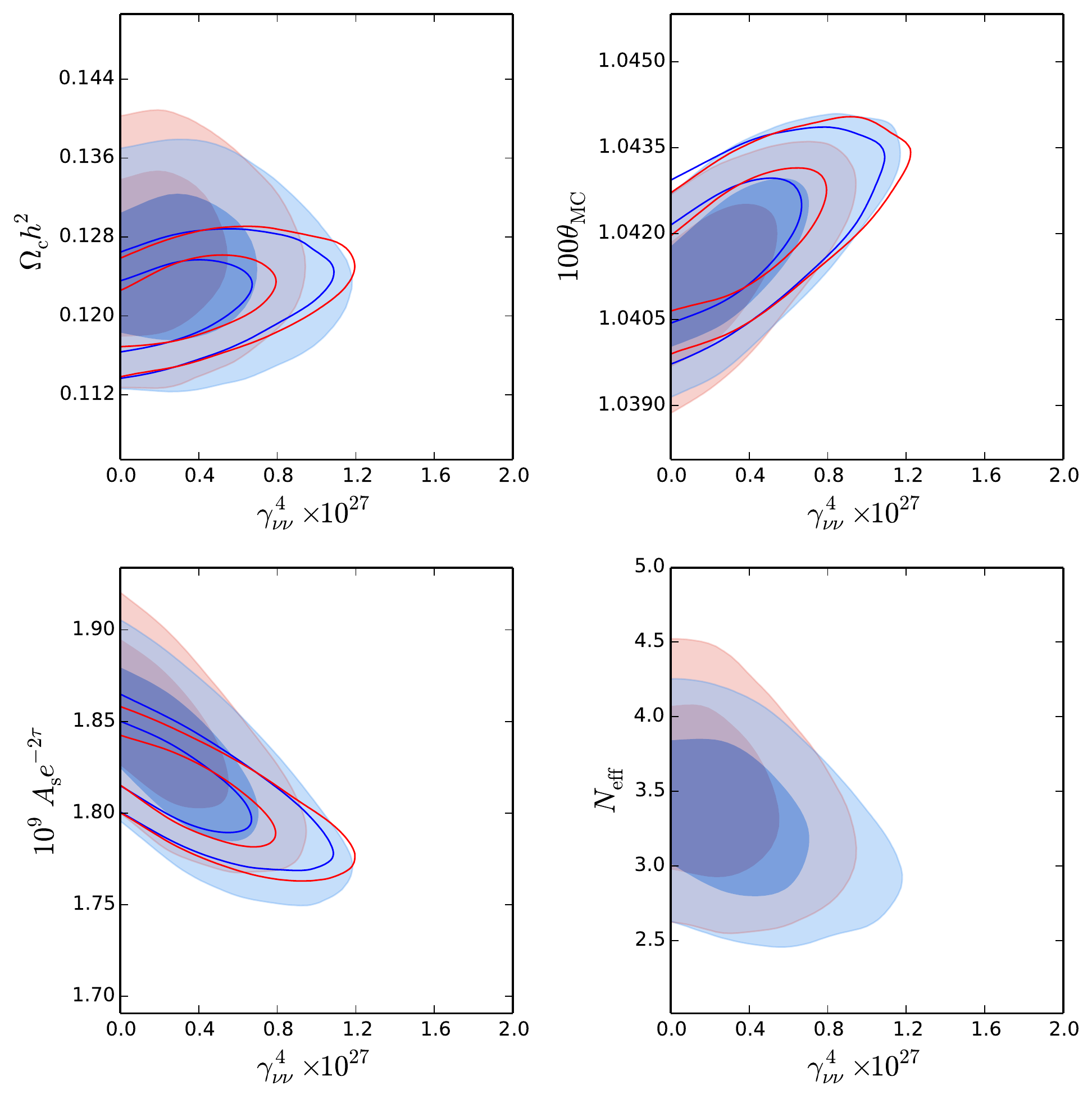}
\caption{68\% and 95\% confidence regions for selected parameter pairs involving $\gnn^4$ 
in the $\Lambda$CDM+$\gnn$ (empty contours) and $\Lambda$CDM+$\gnn+\neff$ (filled contours),
for Planck+WP (blue) and Planck+WP+HL (red).}
\label{fig:gnn_post2D}
\end{center}
\end{figure*}

\begin{figure*}
\begin{center}
\includegraphics[width=1.\textwidth,keepaspectratio]{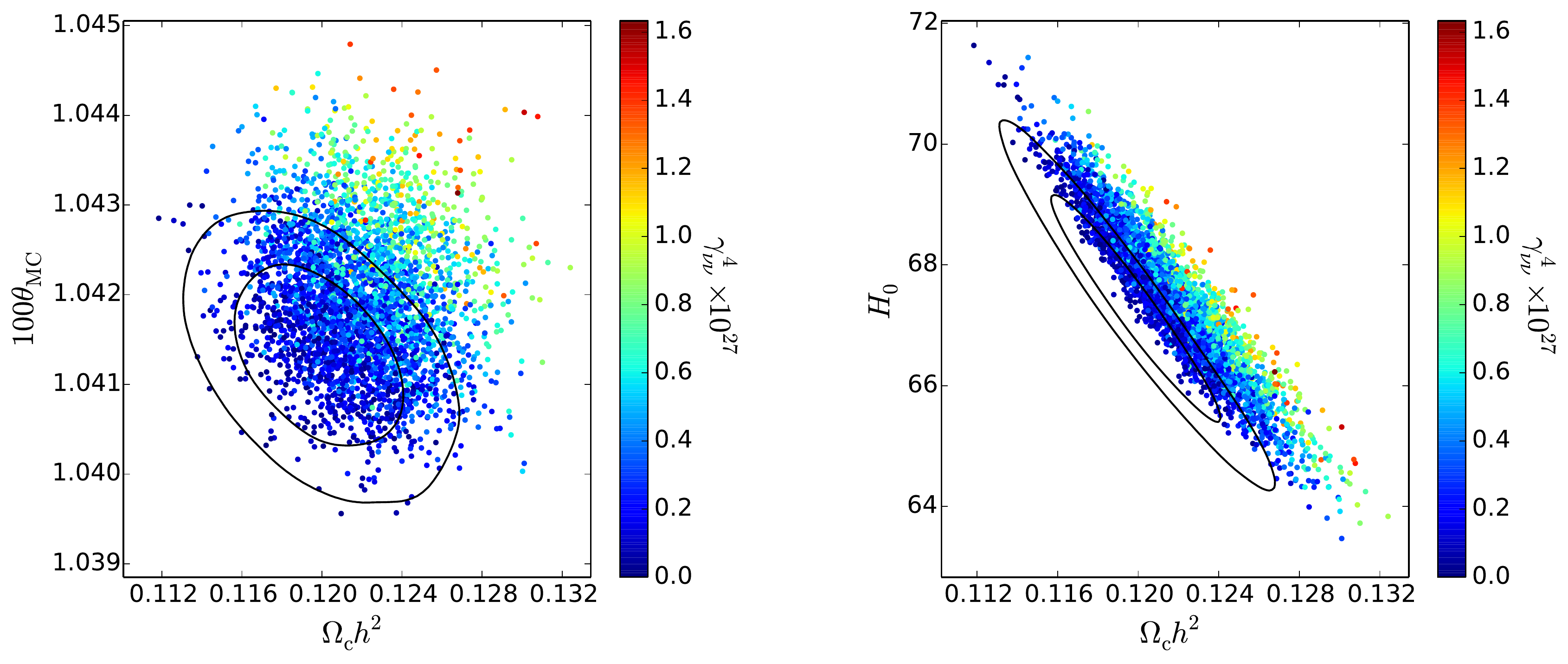}
\caption{Left panel: Samples from Planck+WP chains in the $\Omega_c h^2 - \theta$ plane, color-coded by $\gnn^4$
for the $\Lambda$CDM+$\gnn$ model. The 68\% and 95\% credible regions for the same dataset in the $\Lambda$CDM model
are also shown (solid lines). Notice
how larger values of the coupling constant for secret interactions require larger values of both $\Omega_c h^2$ and $\theta$, as 
explained in the text. Right panel: The same as the left panel, but in the $\Omega_c h^2 - H_0$ plane.}
\label{fig:post3D}
\end{center}
\end{figure*}
Finally, we have also considered the possibility of a non-vanishing amplitude of tensor modes. We label this model as $\Lambda$CDM+$\gnn$+$r$.
In this case, in addition to Planck+WP, we have also used the joint BICEP2/Planck 2015 (BKP) dataset. At 95\% CL, we find $\gnn^4 < 0.90\times 10^{-27}$ ($\gnn < 1.73\times 10^{-7}$ and $\zrec \lesssim 9300$) for Planck+WP and $\gnn^4 < 0.87\times 10^{-27}$ ($\gnn < 1.72 \times 10^{-7}$ and $\zrec \lesssim 9000$) for Planck+WP+BKP. 
Even in this extension of the $\Lambda$CDM model, cosmological data slightly prefer a non-zero value of the coupling: at 68\% CL, we have $\gnn^4 =(0.408^{+0.137}_{-0.357})\times10^{-27}$ for Planck+WP and  $\gnn^4=(0.395^{+0.134}_{-0.342}) \times 10^{-27}$ for Planck+WP+BKP. The posterior probability for $\gnn^4$ in the two cases is shown in Fig. \ref{fig:gnn_post_tens}.
For what concerns the tensor-to-scalar ratio, we find $r < 0.14$ and $r < 0.10$ for Planck+WP and Planck+WP+BKP, respectively. Both values are consistent with those reported in Refs. \cite{Ade:2013zuv, Ade:2015tva} for the $\Lambda$CDM+$r$ model.

\begin{table}
\begin{small}
\begin{center}
\begin{tabular}{c|c|c}
\hline\hline
 & \multicolumn{2}{c}{}\\
 & \multicolumn{2}{c}{$\Lambda$CDM+$\gamma_{\nu\nu} + r$} \\[0.1cm]
   \cline{2-3}
 &                    &                              \\[0.1cm]
 & Planck+WP          & Planck+WP                    \\
 &                    & +BKP                         \\
Parameter             &                              \\
\hline 
 &                    &                              \\
$\Omega_{\textrm{b}}h^2$&$0.02220\pm0.00029$               &$0.02217\pm0.00029$     \\[0.1cm]
$\Omega_{\textrm{c}}h^2$&$0.1213\pm0.0029$                &$0.1217\pm0.0029$       \\[0.1cm]
100 $\theta$            &$1.04211^{+0.00076}_{-0.00085}$  &$1.04203^{+0.00074}_{-0.00083}$       \\[0.1cm]
$\taureio$              &$0.091^{+0.012}_{-0.014}$        &$0.091^{+0.013}_{-0.014}$       \\[0.1cm]
$n_s$                   &$0.9669\pm0.0078$                &$0.9658\pm0.076$       \\[0.1cm]
$\log[10^{10} A_s]$     &$3.075^{+0.025}_{-0.028} $        &$3.078\pm0.025$         \\[0.1cm]
$10^{27}\gnn^4$       & $< 0.90$ & $< 0.87$	   \\[0.1cm]
$r$                     &   $ < 0.14$                      &     $<0.10$           \\[0.1cm]
\hline
$H_0$ [km/sec/Mpc] & $67.7 \pm 1.2$       & $67.5 \pm 1.2$ \\[0.1cm] 
$10^{7}\gnn$       & $< 1.73$ & $<1.72$	   \\[0.1cm]
$r_{0.002}$  & $< 0.13$                 & $ < 0.09$ \\[0.1cm] 
\hline\hline
\end{tabular}
 \caption{Constraints on cosmological parameters for the $\Lambda$CDM+$\gnn$+$r$ model from the analysis of the Planck+WP and Planck+WP+BKP datasets. We quote 68\% C.L., except for upper bounds, which are $95\%$~C.L. }
 \label{tab:pars2}
 \end{center}
 \end{small}
 \end{table}
 
\begin{figure}
\begin{center}
\includegraphics[scale=0.55]{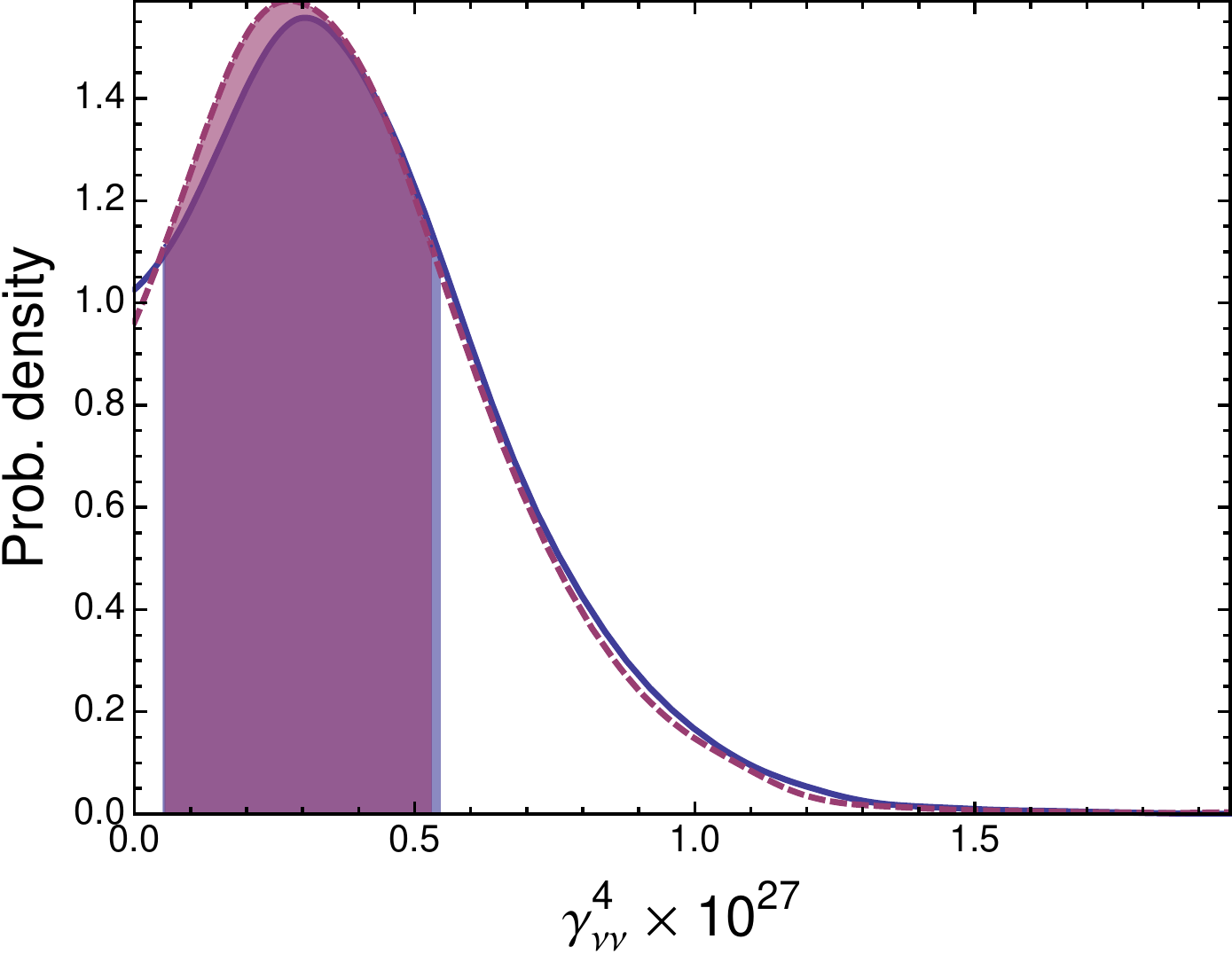}
\caption{One-dimensional posterior distribution for $\gnn^4$ obtained from the Planck+WP (blue solid) and Planck+WP+BKP (red dashed)
datasets, for the $\Lambda$CDM + $\gnn+r$ model. The shaded regions denote the 68\% credible interval.}
\label{fig:gnn_post_tens}
\end{center}
\end{figure}
 
\section{Conclusions}
\label{sec:conc}

We have derived constraints on non-standard neutrino interactions mediated by a (pseudo)scalar massless boson
using observations of CMB temperature and polarization anisotropies from Planck, WMAP, ACT, SPT and BICEP2/KECK.
We have found that, both in a minimal extension of the $\Lambda$CDM model and in more complicated scenarios
allowing for the presence of extra relativistic degrees of freedom or of primordial tensor perturbations, the
strength of non-standard interactions (expressed through the coefficient of the collision term in the Boltzmann equation
for neutrinos) is constrained at 95\% C.L. $\gnn \lesssim 1.7\times 10^{-7}$, quite stable with respect to the models and datasets considered. 
This corresponds to a largest redshift of neutrino recoupling of $\zrec \simeq 8500$, larger than the value $\zrec < 1887$ found in
 Ref. \cite{Archidiacono:2013dua}, and shows that the possibility of neutrino recoupling happening before recombination
 is allowed by the data. On the other hand, we confirm the preference, also reported in Ref. \cite{Archidiacono:2013dua},
 for non-zero values of the coupling constant. We find best-fit values of $\gnn$ in the range $(1.2 \div 1.4)\times 10^{-7}$,
 corresponding to  $\zrec$ in the range $1300 \div 3300$. For comparison, in Ref.  \cite{Archidiacono:2013dua} 
it is found that the probability distribution peaks in $\zrec\simeq 1500$. 
In most cases, we find $\gnn\neq 0$ at 68\% CL; this tendency is more pronounced when small-scale CMB observations, which are sensitive to details
of the photon damping regime, are considered, but is alleviated in presence of extra relativistic degrees of freedom if one allows
for them. On the other hand, considering a non-vanishing amplitude of tensor modes, still leads to a preference for non-zero coupling 
at the same level, even for the base Planck+WP dataset.
 
The exact relationship between our parameter $\gnn$ and the elements of the Yukawa matrix $g_{ij}$ depends on
the details of the underlying particle physics model. As an example, let us consider
the class of models in which neutrino acquire mass through violation of ungauged lepton number.
In this case the neutrino mass eigenstates couple diagonally, to lowest order approximation, 
to the Nambu-Goldstone boson of the broken global symmetry, the Majoron. Neutrino masses are 
proportional to the diagonal couplings: $m_i \propto g_{ii}$.
 Neglecting the small off-diagonal couplings, $g_{ij} = \delta_{ij} g_i$, and further assuming that
the diagonal ones are of the same order of magnitude, $g_i\simeq g$, we have that
\begin{equation}
\Gamma_{\mathrm{bin}} = n_\mathrm{eq} \langle \sigma_\mathrm{bin} v \rangle \simeq 
(1.8 \times 10^{-3})\,g^4 T_\nu,\,
\end{equation}
where we have used $\sigma = g^4/(32\pi s)$  \cite{Kolb:1987qy} for
the neutrino-neutrino scattering cross section and  $n_\mathrm{eq}$ is the abundance of \textit{single} neutrino family.
Comparing this with Eq. (\ref{eq:cambpar_Gamma}) immediately yields $g \lesssim 8.2\times 10^{-7}$. 
This region partially overlaps with the interval $3\times 10^{-7} \lesssim g \lesssim 2\times 10^{-5}$ excluded by observations of SN1987A 
\cite{Kachelriess:2000qc}, although as discussed in Sec. \ref{sec:theory} SN observations do not directly probe
the diagonal elements of the coupling matrix in the mass base. The best-fit values for $\gnn^4$ translate to 
$g \simeq (5 \div 6) \times 10^{-7} $ in the Majoron model, which is also in tension with SN bounds, albeit
the same remark as above applies.

\section{Acknowledgments}
Work supported by ASI through ASI/INAF Agreement I/072/09/0 for the Planck LFI Activity of Phase E2.




\end{document}